\documentstyle[12pt]{article}
\begin{document}
\typeout{ Hi ! This is a Latex file. Please run twice}
\typeout{BLACK HOLES FROM NON-ABELIAN TODA THEORIES}
\typeout{Jean-Loup GERVAIS and Mikhail V.  SAVELIEV}

\begin{titlepage}
\nopagebreak
\begin{flushright}

hepth@xxx/9203039
\\
LPTENS--92/07
\\
ENSLAPP-L-370/92
\\
March  1992
\end{flushright}

\vglue 2.5  true cm
\begin{center}
{\large\bf
BLACK HOLES \\ FROM \\ NON-ABELIAN TODA THEORIES} \\

\vglue 1  true cm
{\bf Jean-Loup~GERVAIS}\\
{\footnotesize Laboratoire de Physique Th\'eorique de
l'\'Ecole Normale Sup\'erieure\footnote{Unit\'e Propre du
Centre National de la Recherche Scientifique,
associ\'ee \`a l'\'Ecole Normale Sup\'erieure et \`a l'Universit\'e
de Paris-Sud.},\\
24 rue Lhomond, 75231 Paris C\'EDEX 05, ~France.
\\
and
} \\
{\bf Mikhail V.  SAVELIEV}\footnote{ On leave of absence from
the Institute for High Energy Physics,
142284, Protvino, Moscow region, Russia.}\\
{\footnotesize Laboratoire de Physique Th\'eorique de
l'ENS Lyon, \\ 46 All\'ee  d'Italie, 69364 Lyon C\'EDEX 07, France.}\\
\medskip
\end{center}

\vfill
\begin{abstract}
\baselineskip .4 true cm
\noindent
Non-abelian Toda theories are shown to provide exactly
solvable conformal systems in the presence of a
black hole. They correspond to gauged WZNW models where the
gauge group is nilpotent, and are thus basically different
  from the ones
currently considered, following Witten. The non-abelian
Toda potential  gives a cosmological term which may be integrated
exactly at the classical level.

\end{abstract}
\vfill
\end{titlepage}
\baselineskip .5 true cm

{}From the beginning the conformal field theories in two-dimensions
(2D CFT) have been considered using  two different viewpoints. The
celebrated
BPZ approach \cite{BPZ},  made a systematic   use of Ward identities
assuming that
renormalization gives   simple numerical  modifications
 of  the coupling constants. It came out
 at a   time where  the authors of ref.\cite{GN} had already
started to solve the
Liouville theory. The latter work  was the beginning of
 the modern investigations
based on the connection between 2D CFT and integrable systems.
In particular, the authors of papers \cite{BG} presented a
classification of the CFT  associated  with the Toda systems,
and of the related W-symmetries by putting them in
correspondence with the simple finite-dimensional Lie algebras
${\cal G}$ over ${\bf C}$.  In the present letter, we discuss
 the so-called non-abelian Toda theories.
Most of  these  models  are derivable from   $\sigma$-model
actions of the form ($\gamma$ is the coupling constant)
\begin{equation}
{\cal S} =\frac{1}{\gamma} \int d_2z \sqrt{-h}\left \{
{1\over 2} G_{ij} (x) h^{\mu \nu}
\partial_{\mu}x^{i}
\partial_{\nu}x^{j} - T(x) +\Phi(x)R^{(2)}\right \}.
\label{1}
\end{equation}
Greek indices are related to the world-sheet which has metric tensor
$h^{\mu \nu}$, and scalar curvature $R^{(2)}$.  Latin indices
run from $1$ to $N$. The functions
 $x^i(z^+, z^-)$ are bosonic fields
  which may be regarded
as components of a string-position in a $N$-dimensional
target space with metric
 $G_{ij} (x)$.  From this viewpoint,
 $T$ and $\Phi$ are the tachyon and
dilaton fields respectively. In the present article, we shall
mostly work with constant  Minkowski 2D metric
$h_{\mu \nu}=-\eta_{\mu \nu}$, in which case, the dilaton field
only appears in the expression of the (improved ) stress-energy
tensor
\begin{equation}
\label{2}
\theta_{\mu \nu}\equiv  2{\delta S \over  \delta h^{\mu \nu}}
\Bigl |_{ h^{\mu \nu} =-\eta^{\mu \nu}}
=T_{\mu \nu}+2(\eta_{\mu \nu} \partial^2 -
\partial_\mu \partial_\nu) \Phi,
\end{equation}
where $T_{\mu \nu}$ is the  canonical (Noether) stress-energy tensor
with the flat 2D metric.
  As an example, the  abelian Toda theories, considered
in ref.\cite{BG} are associated with the Lagrangian
\begin{equation}
\label{3}
-\gamma {\cal L} = \frac{1}{2}{\sum^{N}_{i,j=1}}
(wK^{({\cal G})})_{ij} {{\partial
x^{i}} \over
{\partial z_{+}}}
{{\partial x^{j}} \over {\partial z_{-}}} + \sum_{i=1}^N
\prod_{j=1}^N  w_{i} \exp \left(K^{({\cal G})}_{ij} x^j\right ),
\end{equation}
where $G_{ij}$ coincides with
the symmetrized Cartan matrix of a simple Lie algebra ${\cal G}$
of rank $N$.  Explicitly, denoting by $\vec \alpha_i$ a set
of primitive roots of ${\cal G}$,
$K^{({\cal G})}_{ij}=(\vec\alpha_i .\vec\alpha_j)2/\alpha_j^2$
is the
Cartan matrix, and
$w  \equiv \mbox{diag}   ( w_{1}, \ldots , w_{N})$,
$w_{i}=2/\alpha_i^2$. These theories are of the type Eq.\ref{1}
with
\begin{equation}
\label{4}
T\equiv \sum_{i=1}^N
\prod_{j=1}^N  w_{i} \exp \left(K^{({\cal G})}_{ij} x^j\right ),
\quad \Phi=\sum_{i=1}^N \omega_i x^i.
\end{equation}
Clearly, the coefficients $\omega_i$ in $T$ may be changed to any
value by shift of the fields $x$. These defines  $N$ cosmological
constants which may be re-introduced whenever needed.
The expression for $\Phi$ follows from the study\cite{BG} of the
conformal invariance properties  of these systems.
The  fields $x^i$  satisfy the equation
\begin{equation}
{\partial{^2}x{^i}} / {\partial z{_+}}
{\partial z{_-}} = \prod_j \exp
\left(K^{({\cal G})}_{ij} x^j\right ),
\quad  1\leq i \leq N;
\label{5}
\end{equation}
It is natural to characterize such systems as being  abelian since,
in the
framework of an algebraic approach,
(see e.g. review \cite{LS1}), they are
associated with the canonical gradation
${\cal G} = \oplus_{m \epsilon
{\bf Z}} {\cal G}_m$ of the simple
Lie algebras, for which the subalgebra ${\cal G}_{0}$ is abelian.
Indeed, the canonical gradation is such that the step operators
asociated with positive (resp. negative) simple roots have grading
$1$ (resp. $-1$), and ${\cal G}_{0}$ coincides with the Cartan
subalgebra. In other words,
 this   canonical gradation  is related with
the  principal $ sl(2, {\bf C})$ subalgebra of ${\cal G}$.

The above example is somewhat trivial from the target-space
viewpoint since the metric $G^{ij}$ is constant. The point of the
present letter is to remove  this restriction, by considering
 the so-called non-abelian versions of the
Toda systems. These theories,  constructed in \cite{LS2},
 correspond to  non-canonical gradations of simple Lie algebras
where  the subalgebra ${\cal G}_{0}$
is non-abelian. Such {\bf Z}--gradations of
${\cal G}$ are related with the
non-principal embeddings of $sl(2,{\bf C})$ in
${\cal G}$. We shall put forward   the  examples  of the Lie
algebras  $B_{N-1}$, since   they
give integrable theories with   black-hole background metrics
similar to the one recently  popularized by
Witten\cite {EFR}\cite{BCR} \cite{RSS} \cite{MSW} \cite{W}.
Associated   with  a  maximal  subalgebra  $D_{N-1}$  (of
maximal rank) in the Lie algebra $B_{N-1}$,
 we have the Lagrangian
\begin{eqnarray}
-\gamma {\cal L} =
\frac{1}{2}\Bigl[\sum_{j ,k =1}^{N-1} K_{jk}^{(D_{N-1})}
\frac{\partial x^{j}}{\partial z_+}
\frac{\partial x^{k}}{\partial z_-}
& - & \tanh^2 \left ({x^{N-1} - x^{N-2}\over 2}\right  )
\frac{\partial x^N}{\partial z_+}
\frac{\partial x^N}{\partial z_-}\Bigr] \nonumber\\
& + & \sum_{i=1}^{N-1}
\prod_{j=1}^{N-1} \exp \left(K^{( D_{N-1})}_{ij} x^j\right ).
\label{6}
\end{eqnarray}
Note that  there  is
no useful non-abelian version of the Toda lattice for the series
$A_N$. In this case, the zero-grading part
is given by $ {\cal G}_0= gl(1)\oplus \cdots \oplus gl(1)
\oplus B_1$,
where the one-dimensional linear algebra
 $gl(1)$ appears $N-1$ times.
The $x$-dependent  part of the metric involves  the
 hyperbolic-tangent-square function  which is familiar
in the 2D black-hole game. In particular if we choose
$N=3$, we  find
\begin{equation}
\label{7}
- \gamma {\cal L} = \frac{1}{2}\left[2\sum_{i =1}^{2}
\frac{\partial x^i}{\partial z_+}
\frac{\partial x^i}{\partial z_-}
 -  \tanh^2 \left ({x^{1} - x^{2}\over 2}\right)
\frac{\partial x^3}{\partial z_+}
\frac{\partial x^3}{\partial z_-}\right]
 +  \sum_{i =1}^{2} \exp (2x^i)
\end{equation}
In order to clarify the black-hole aspect,
it is convenient to change field variables. Let
\begin{equation}
\label{8}
\phi=(x^1+x^2)/2,\quad
r=(x^1-x^2)/2, \quad
\theta=x^3/2.
\end{equation}
One gets
\begin{equation}
\label{9}
-{\gamma \over 2} {\cal L} =
\frac{\partial \phi}{\partial z_+}
\frac{\partial \phi}{\partial z_-}
+\frac{\partial r}{\partial z_+}
\frac{\partial r}{\partial z_-}
 -  \tanh^2 \left (r\right)
\frac{\partial \theta}{\partial z_+}
\frac{\partial \theta}{\partial z_-} \nonumber\\
 +  \cosh (2r) e^{2\phi}.
\end{equation}
The target-space metric is
\begin{equation}
\label{10}
G =\left (\begin{array}{ccc}
1&0&0\nonumber\\
0&1&0\nonumber \\
0&0&-\tanh^2 r
\end{array}\right )
\end{equation}
one sees that
the first component,   which corresponds to the $\phi$ variable,
defines a subspace that
 decouples from the rest
from the viewpoint of Riemannian geometry.
In the ($r$, $\theta$)  space we have
$G=\left (\begin{array}{cc}
1&0\nonumber \\
0&-\tanh^2 r
\end{array}\right ),
$
which coincides  with Witten's black-hole metric\cite{W} exactly.
As is well known by now the corresponding manifold is a cigar
with radial variable $r$, and angle $\theta$. Letting
\begin{equation}
\label{11}
u=\sinh (r) e^{-\theta},\quad
v=-\sinh (r) e^{\theta},
\end{equation}
one finds the standard form $
-du dv \bigl / (1-uv)=dr^2- d\theta^2\tanh ^2 r $.
The action becomes
\begin{equation}
\label{12}
{\gamma\over 2} {\cal L}=
{1\over (1-uv)} \bigl (\partial_+u \partial_-v
+\partial_-u \partial_+v\bigr ) -\partial_+\phi \partial_-\phi
-e^{2\phi}(1-2uv).
\end{equation}
Note that there exists another non-canonical gradation of $B_N$,
with ${\cal G}_0=gl(1)\oplus B_{N-1}$, for which the corresponding
non-abelian version of the Toda systems possess the
Lagrangian
\begin{equation}
\label{12a}
{\gamma \over 2} {\cal L}=
{1\over (1-\vec u.\vec v)} \bigl
(\partial_+\vec u.  \partial_-\vec v
+\partial_-\vec u. \partial_+\vec v\bigr ) -
\partial_+\phi \partial_-\phi
-e^{2\phi}(1-2\vec u.\vec v),
\end{equation}
where $\vec u$ and $\vec v$ are $(N-1)$-dimensional vectors.
It  represents  a  black higher-dimensional manifold.

For completeness, let us next recall some standard results on the
systems we are discussing.
The basic point of the Toda theories --- abelian as well as
 non-abelian --- is that
their dynamical equations, no matter how complicated, are derivable
from  flatness  conditions
\begin{equation}
\label{13}
\Bigl[\partial_+-{\cal A}_+,
\partial_--{\cal A}_-\Bigr]\equiv \partial_-{\cal A}_+
-\partial_+{\cal A}_-+\Bigl[ {\cal A}_+,{\cal A}_- \Bigr]=0,
\end{equation}
 which
allow  to obtain the general solutions in closed
form\cite{LS1}\cite{LS2}. The Lax pair
${\cal A}_\pm$ is systematically constructed once the
Lie algebra, the gradation, and the grading spectrum
of ${\cal A}_\pm$  are chosen\cite{LS1}\cite{LS2}.
These results were re-derived recently\cite{Dublin}
by proving that the Toda theories may be obtained by
conformal reduction
of  WZNW models. Let us  recall some basic
point about Toda solutions, following refs.\cite{LS1}\cite{LS2},
but stressing the WZNW connection of ref.\cite{Dublin}.
Consider a finite-dimensional Lie group ${\bf G}$,
with Lie algebra $\cal G$, and a gradation
${\cal G} = \oplus_{m \in
{\bf Z}} {\cal G}_m$, with  $[{\cal G}_m, {\cal G}_n ]
 \in {\cal G}_{m+n}$. There is a grading operator $H$, such that
$[H, {\cal G}_m ] =
 2m  {\cal G}_{m}$. Associated with this grading, we have the
(so-called modified) Gauss
decomposition of the Lie group ${\bf G}$, namely, any
regular element $\omega$ may be
written as
\begin{equation}
\label{14}
\omega\equiv \omega_+\omega_0\omega_-, \quad \hbox{with}\>
\omega_\epsilon \in {\bf G}_\epsilon, \quad \epsilon = \pm, \, 0.
\end{equation}
 ${\bf G}_0$, ${\bf G}_\pm$ are the subgroups generated by
${\cal G}_0$, and
${\cal G}_{\pm}\equiv  \oplus_{m >0}
 {\cal G}_{\pm m} $, respectively.
With the simplest choice of gradation spectrum,
the Lax pair is given by\footnote{By convention, the differential
operators $\partial_\pm$ only act on the first function on the right,
unless parenthesis indicate otherwise.}
\begin{eqnarray}
\label{15}
{\cal A}_-&\equiv &  \partial_-g_0^{-1} g_0-M_-, \nonumber\\
{\cal A}_+&\equiv &  -g_0^{-1} M_+ g_0,
\end{eqnarray}
where $g_0\in {\bf G}_0$, and
where $M_\pm$ are, respectively,
fixed elements of ${\cal G}_{\pm 1}$,
such that the action of ${\cal G}_0$ on them generates  the whole
${\cal G}_{\pm 1}$. The connection with the WZNW model goes as
follows\cite {Dublin}. Let $\omega_0=g_0^{-1}$,
and define  $\omega_{\pm}$ as solutions
of the equations
\begin{equation}
\label{16}
\omega_+^{-1} \partial_+ \omega_+ =\omega_0 M_+\omega_0^{-1},
\quad
\omega_- \partial_- \omega_-^{-1} =\omega_0^{-1} M_-\omega_0.
\end{equation}
Clearly, the solutions are such that
$\omega_\pm \in {\bf G}_\pm$, and if we
define $\omega\equiv \omega_+\omega_0\omega_-$, we have a
Gauss decomposition of the type  Eq.\ref{14}.
A straightforward calculation
then shows\cite{Dublin}  that  the currents
\begin{equation}
\label{17}
\partial_- \omega \> \omega^{-1}=J,\quad
\omega^{-1} \partial_+ \omega=-\bar J.
\end{equation}
satisfy the equations
\begin{equation}
\label{18}
\partial_+J=\partial_-\bar J=0,
\end{equation}
and it follows that $\omega$ is a solution of  the WZNW equations.
Moreover, one may verify that the currents are such that their
projections $\Pi_\pm$ into ${\cal G}_\pm$ are given by
\begin{equation}
\label{19}
\Pi_-\bigl (J\bigr )=-M_-, \quad
\Pi_+\bigl (\bar J\bigr )=-M_+.
\end{equation}
Next, the general solution  of the dynamical equation asssociated
with Eq.\ref{15}, that is
\begin{equation}
\label{20}
\partial_+\left ( \partial_-  g_0^{-1} g_0 \right )
+\left [ M_- ,  g_0^{-1} M_+ g_0\right ] =0
\end{equation}
has been known for quite some time\cite{LS1}\cite{LS2}\cite{LS3}.
It was recently
connected\cite{Dublin}
with the WZNW solution  as follows. The general solution of
Eq.\ref{18} is $\omega=\omega_L(z_-) \omega_R(z_+)$ with
$J=\partial_- \omega_L \omega_L^{-1}$,
$\bar J=-\omega_R^{-1}\partial_+ \omega_R^{-1}$.
 The group
element $g$ relevent for the Toda theory is $g\equiv \omega^{-1}$,
and we have
$g=g_L(z_+) g_R(z_-)$, with $g_L=\omega_R^{-1}$, $g_R=\omega_L^{-1}$
and $g= g_- g_0 g_+$, $g_\pm \equiv \omega_\pm^{-1}$.
It next follows from
Eqs.\ref{16} that, if we introduce the decompositions
$g_L=g_{-L}\, g_{0L}\, g_{+L}$, $g_R=g_{-R}\, g_{0R}\, g_{+R}$,
we have
\begin{eqnarray}
\label{21}
\partial_+ g_{+L}& =-\bar f(z_+)\>  g_{+L},\quad
\bar f(z_+) &\equiv g_{0L}^{-1} M_+ g_{0L} \in {\cal G}_+ \nonumber \\
\partial_- g_{-R} &= g_{-R}\> f(z_-),\quad
f(z_-) &\equiv g_{0R}^{-1} M_- g_{0R} \in {\cal G}_-,
\end{eqnarray}
with solutions
\begin{eqnarray}
\label{22}
g_{+L}(z)&=& \sum_n (-1)^n \int ^z dx_1 \cdots \int^{x_{n-1}} dx_n
\> \bar f(x_1) \cdots \bar f(x_n),  \nonumber \\
g_{-R}(z)&=& \sum_n \int ^z dx_1 \cdots \int^{x_{n-1}} dx_n
\> f(x_n) \cdots f(x_1),
\end{eqnarray}
where only a finite number of terms are non-vanishing, since the
Lie algebra considered is finite-dimensional. The solution
of  Eq.\ref{20} is obtained by computing the matrix-elements of $g$
between highest-weight vectors $| \lambda >$ which are annihilated
by ${\cal G}_m$ for positive $m$, so that $<\lambda | g | \lambda >
=<\lambda | g_0 | \lambda >$. One finds
\begin{equation}
\label{23}
<\lambda | g_0 | \lambda >=
<\lambda | g_{L0} g_{L+} g_{R-}g_{R0} | \lambda >,
\end{equation}
which, according to Eq.\ref{22}, coincides with the general solution
of refs.\cite{LS1}\cite{LS2}\cite{LS3}. In practice, $f$ and $\bar f$
are arbitrary elements of ${\cal G}_-$ and  ${\cal G}_+$,
respectively,  functions
of a single variable. In the same way as for
the abelian case\cite{BG},
they are classical analogues  of the quantum screening
operators.  In general, the zero-grading part
${\cal G}_0$ involves a subalgebra ${\cal G}_0^0$, that commutes
with $M_\pm$. It follows from Eq.\ref{20}
 that in $g_0$, only
the part that do not belong to ${\cal G}_0^0$ has a non-trivial
dynamics.  This terminates the short summary of general properties.

Returning to the case of the non-abelian  $B_2$ Toda theory,
it is illutrative to make use of  the following  explicit
realization\cite{G}  of the Lie algebra $B_2$.
 Consider two orthonormal vectors  $\vec e_{1,2}$
in an Euclidian 2D vector space.
The positive roots of $B_2$ are
\begin{equation}
\label{24}
\vec \pi_1=\vec e_1-\vec e_2,\>
\vec \pi_2=\vec e_2,\> \vec \pi_{12}=\vec e_1,
\vec \pi_{122}=\vec e_1+\vec e_2,
\end{equation}
where the first two are simple. Introduce five  fermionic
oscillators $\flat_j$, $j=\pm 1$, $\pm 2$, $0$,
with $[\flat_j, \flat_k^+]=\delta_{j,k}$. The
$B_2$   generators are
\begin{eqnarray}
\label{25}
h_1&=&\flat_1^+ \flat _{1} -\flat_{-1}^+ \flat _{-1}
-\flat_2^+ \flat _{2} -\flat_{-2}^+ \flat _{-2}, \nonumber \\
h_2&=&2(\flat_2^+ \flat _{2} -\flat_{-2}^+ \flat _{-2})\nonumber \\
E_{e_1- e_2}&=&\flat_1^+ \flat _{2}
-\flat^+_{-2} \flat_{-1}, \nonumber \\
E_{e_2}&=&\sqrt{2}(\flat_2^+ \flat _{0}
-\flat^+_{0} \flat_{-2}),  \nonumber \\
E_{e_1}& =&\sqrt{2}(\flat_1^+ \flat _{0}
-\flat^+_{0} \flat_{-1}),  \nonumber \\
E_{e_1+ e_2}&=&\flat_1^+ \flat _{-2}
-\flat^+_{2} \flat_{-1}.
\end{eqnarray}
The non-abelian Toda theory arises by choosing
\begin{equation}
\label{26}
H=2(\flat_1^+ \flat _{1} -\flat_{-1}^+ \flat _{-1}), \quad
M_{+}=E_{e_1}, \quad M_{-}=E_{e_1}^+.
\end{equation}
These operators generate a non-canonical
embedding of $A_1$ in $B_2$.
The gradation $0$ part ${\cal G}_0$ is the linear span of
$h_1,\, h_2,\, E_{e_2}, E_{e_2}^+$,
and ${\cal G}_0^0$ is clearly
generated
by $h_2$.  The subspace ${\cal G}_1$ is the linear span of
 $E_{e_1-e_2}, \, E_{e_1+e_2}, \,
E_{e_1}$. The first two are step operators corresponding to simple
roots of $D_2$. The parametrization of $g_0$ is
\begin{equation}
\label{27}
g_0= \exp (a^+E_{e_2}) \exp (a^-E_{e_2}^+) \exp(a_1 h_1+a_2h_2),
\end{equation}
which gives
\begin{eqnarray}
\label{28}
g_0^{-1}\partial_-g_0 &=& \partial_-a_1H/2 +
(\partial_-a_2-\partial_-a_1/2  + \partial_-a^{+}a^-)h_2
+e^{-2a_2+a_1}\partial_-a^{+}
E_{e_2} + \nonumber\\
&+& e^{2a_2-a_1} [\partial_-a^{-} - \partial_-a^{+}(a^-)^2]E_{e_2}^+.
\end{eqnarray}
\begin{eqnarray}
\label{29}
g_0^{-1}M_+g_0 =  e^{-a_1}\Bigl  \{-2a^+e^{a_1-2a_2}E_{e_1+e_2}+
(1+2a^+a^-)E_{e_1}+\nonumber \\
   2a^-(1+a^+a^-)e^{-a_1+2a_2}E_{e_1-e_2} \Bigr \}.
\end{eqnarray}
The final equations, equivalent to Eq.\ref{20} are
\begin{eqnarray}
\label{30}
\partial_{+}\partial_-a_{1} & = & -2(1+2a^+a^-) e^{-a_1},\nonumber \\
\partial_{+}\partial_-a_{2} +\partial_+
\left(\partial_-a^+a^-\right )
 & = & -(1+2a^+a^-) e^{-a_1},
\nonumber\\
\partial_+\left \{e^{a_1 - 2a_2} \partial_-a^+\right \}
&=& 2a^+e^{-2a_2},\nonumber\\
\partial_+\left \{e^{-a_1 + 2a_2}
\partial_+\left (-\partial_-a^+(a^-)^2+\partial_-a^{-}\right)
\right\}&=&
 2a^-(1+a^+a^-)e^{-2a_1 +2a_2}.
\end{eqnarray}
These equations are equivalent  to the field-equations  derived
from the action Eq.\ref{9} if we let
\begin{equation}
\label{31}
\phi=-a_1,\quad \sqrt{a^+a^-} =\cosh (r/2),
\end{equation}
\begin{eqnarray}
\label{32}
\partial_+\theta/\sqrt{2}&=& -(1+a^+a^-)
\partial_+\left \{ \ln\left (a^- e^{-a_1+2a_2}\right ) \right \}
+\partial_+\left \{ \ln a^+a^-\right \}/2, \nonumber \\
\partial_-\theta/\sqrt{2}&=& {1+a^+a^-\over 1+2a^+a^-}
\partial_-\left \{ \ln\left (a^+ e^{a_1-2a_2}\right ) \right \}
-\partial_-\left \{ \ln a^+a^-\right \}/2.
\end{eqnarray}
Finally, one makes use of Eq.\ref{23} with the three possible
highest-weight states
\begin{equation}
\label{33}
\vert \lambda_1 >= \flat_1^+ \vert 0 >, \>
\vert \lambda_2 >= \flat_2^+ \flat_1^+\vert 0 >, \>
\vert \lambda_0 >= \flat_0^+ \flat_1^+\vert 0 >.
\end{equation}
Combining the result  with Eq.\ref{21} gives the exact solutions
in terms of the screening fields $f$ and $\bar f$ (see
ref.\cite{LS2} for details).  Such is the
classical Coulomb gas picture of the black-hole solution.

Our next topic is the connection
with gauged WZNW models. Witten's solution is obtained
from the $sl(2,R)$ -- WZNW model by gauging the $gl(1,R)$ subgroup
$\omega\to \exp (\alpha \sigma_3)\>  \omega \exp (\alpha \sigma_3)$,
 $\alpha$ being an  arbitrary function. It may be generalized
straightforwardly by embedding this $sl(2,R)$ into larger groups.
Its basic characteristic is that one gauges a diagonal
one-parameter subgroup.
 In our explicit example, the non-abelian
Toda theory was based on the   Lie algebra
$B_{N-1}$. Then the group of the corresponding WZNW model is
 the associated maximally-non-compact real form,
that is,  $so(N-1,N)$. As shown in ref.\cite{Dublin},
the Toda theories (abelian or not) are obtained by gauging
a nilpotent algebra instead of a diagonal one, and
the gauge-group transformation   is given by
$\omega \to h_+ \omega h_-^{-1}$,
where $h_\pm \in {\bf G}_\pm$,  and  ${\bf G}_\pm$
is specified by the gradation of
the Lie algebra considered (see above). For instance, for $N=2$
the action Eq.\ref{9} is obtained from the $so(1,2)$
-- WZNW model by  gauging the  transformation
\begin{equation}
\label{34}
\omega \to \exp( \alpha E_{e_1-e_2}+\beta E_{e_1+e_2}
+\gamma E_{e_1}) \>\omega\>
\exp( -\tilde \alpha E_{e_1-e_2}^+-\tilde \beta E_{e_1+e_2}^+
-\tilde \gamma E_{e_1}^+),
\end{equation}
where $\alpha, \cdots, \, \tilde \gamma$ are arbitrary.
Clearly this mechanism is more complicated than Witten's.
However the dynamics
is completely integrable, and we are able to obtain the solution in
closed form to arbitrary order in the tachyon field
$T\propto e^{2\phi} (1-2uv)$. Thus this particular tachyon field
 is exactly solved.

Now  we come to the dilaton field $\Phi$, and to the
conformal invariance. At the purely classical   level
(following ref.\cite{BG}), $\Phi$ is determined from the
requirement that $\theta_{\mu \nu}$
--- which is defined by Eq.\ref{2} ---
be traceless, as a consequence of the equations  of motions
\begin{equation}
\label{35}
{\partial^{2}x^{i}} /{\partial z_{+} {\partial z_{-}}} -
\sum_{j}G^{ij} (x)
{{\delta T(x)} \over {\delta x^{j}}}+\sum_{j,l} \Gamma_{jl}^{i}(x)
{{\partial x^{j}} \over {\partial z_{+}}}
{{\partial x^{l}}\over{\partial z_{-}}} =0.
\end{equation}
Here   $\Gamma_{jl}^{i}(x)$ are the Christoffel symbols
associated with $G$.  Conformal invariance requires that
\begin{equation}
\label{36}
{{\partial^{2}\Phi(x)} / {\partial z_{+} \partial z_{-}}}-T(x)=0,
\end{equation}
which is equivalent to
$$
\sum_{i,j} \left (
\partial_+x^i \partial_- x^j \Phi_{;ij}
+G^{ij} {{\delta \Phi} \over {\delta x^{i}}}
{{\delta T} \over {\delta x^{j}}}\right )  - T = 0
$$
where semi-colons denote covariant derivatives with respect to
$x$. For a given value of $z_\pm$, $\partial_\pm x^j$, and
$x^j$ are independant vectors. Since $T$ and $\Phi$ are taken
to be independant of $\partial_\pm x^j$, it follows that we
must have
\begin{equation}
\label{37}
\Phi_{;ij}=0, \> i,j=1,\cdots, N;\quad
\sum_{\ell,m}G^{\ell m } {{\delta \Phi} \over {\delta x^{\ell}}}
{{\delta T} \over {\delta x^{m}}}  - T = 0.
\end{equation}
The equations
$v_{i;j}  \equiv  \partial  v_i  /   \partial   \phi^{j}   -
\Gamma_{ij}^l v_{l}=0$
defines  the field of  parallel vector fields
 $v_{i}(x)$. The
integrability conditions for these equations, that is,
 the existence of such
parallel vectors  was established in ref.\cite{ES}.
The answer is that it should be
possible to choose a  coordinate system, in which the aforementioned
conditions can be
expressed as decomposability of the matrix
$G_{ij}$, $1\leq i,j\leq N$,
into a direct sum of a constant diagonalizable matrix
$G_{i j}^{(0)}$,
$1\leq i, j\leq s $, and a matrix
$G_{ij}^{(1)}$, $s+1\leq i,j\leq N$,
depending only on the coordinates $x^{s +1},\ldots,x^{N}$,
that is,
\begin{equation}
\label{38}
G=
\left ( \begin{array}{cccccc}
G_{11}^{(0)} & \cdots & G_{1s}^{(0)} & 0 & \cdots & 0
\nonumber \\
\vdots    & \vdots    &\vdots    &\vdots    &\vdots    & \vdots
\nonumber \\
G_{ s 1 }^{(0)} & \cdots & G_{s s}^{(0)}& 0
& \cdots & 0 \nonumber \\
 & & & & &
\nonumber \\
0     & \cdots & 0 &  G_{s+1 \, s+1}^{(1)}& \cdots
&G_{s+1 \, N}^{(1)}
\nonumber \\
\vdots    & \vdots    &\vdots    &\vdots    &\vdots    & \vdots
\nonumber \\
0     & \cdots & 0 &  G_{N \, s+1}^{(1)}& \cdots
&G_{N \, N}^{(1)}
\end{array} \right ).
\end{equation}
The  integer $s$,  which  is between $1$ and $N$,  is
the number of linearly independent vectors
($v_{i}^{(1)},\ldots, v_{i}^{(s)}$)
which generate a complete set of solutions of  the system
$$
\sum_{n}     R^{l} _{nij} v^{n} =0,     \qquad
\sum_{n}     R^{l} _{nij; m_{1} \ldots m_{q}}
v^{n}=0,                                 \qquad
q \geq 0;
$$
$R^{l}_{nij}$ are the components of the Riemann tensor. (Note that
the
algebraic compatibility of these equations testifies to the existence
of the parallel vector fields, for more details see ref.\cite{E}.)
With these coordinates, the covariantly constant vectors
$v^{(\ell)}_i$ are
actually constant, and $v_i^{(\ell)}=0$ for $s+1 \leq i \leq N$. Thus
the covariantly constant vector $\Phi_{;j}
\equiv \delta \Phi/\delta x^{j}$
satisfies
\begin{equation}
\label{39}
\delta \Phi/\delta x^{i}= \left \{\begin{array}{ccc}
 c_i,  & \> \hbox{if} \> &1\leq i\leq s \nonumber \\
0,   & \> \hbox{if} \> &s+1\leq i\leq N, \end{array} \right.
\end{equation}
where the $c_i$ are constants. With  those coordinates,
$\Phi=\sum_1^s c_i \phi^i$ and Eq.\ref{36} becomes
($c^j= G^{(0)\, ji}c_i$)
\begin{equation}
\label{40}
\sum_{i =1}^s c^j \partial T \bigl / \partial x^j
-T=0.
\end{equation}
The general solution of this equation is
$T=\exp\left(\Phi/\sum_{i} c_i c^i\right) \tilde T$,
where $\sum_i c^j \partial \tilde T \bigl / \partial x^j=0$.
One may verify  that this equation is satisfied by the Toda
theories\cite{BG}, as well as by their non-abelian versions.
In particular  our black-hole solution is conformally
invariant, at the
classical level.  However, it is easy to see that
$e^{2\phi}$ transforms  as a $(1,1)$  operator
while $r$ and $\theta$ have
vanishing weights. Thus the background metric Eq.\ref{10} is
trivially invariant.  More generally, Eq.\ref{39} shows that
the dilaton field does not depend upon the variables
$x^s, \cdots , x^N$ for which the background metric is non-trivial.
Thus the latter is classically inert under conformal transformation.
Clearly $T$ plays the role of the 2D cosmological term.

Our next topic is the $W$ algebraic structure of our black-hole
solutions. As recalled at the beginning, the abelian Toda systems
provide\cite{BG} Noether realizations of the W-algebras. It is
interesting to note that for the non-abelian versions of the
Toda systems, the W-algebras are basically different. In particular,
for the case of $B_2$, all  three elements  are of
second oder\cite{S1}. They may be written as (we consider one of the
two holomorphic components,  say, the $z_+$ one).
\begin{equation}
\label{41}
W_{\pm}= 2 p_\pm p_0-\partial_+p_\pm, \quad
W_0=p_0^2 +p_+p_--\partial_+ p_0
\end{equation}
where the fields $p_\pm$, $p_0$ are expressed as
\begin{equation}
\label{42}
p_\pm =e^{\pm \nu} (\pm \tanh (r) \partial_+ \theta +\partial_+ r),
\quad p_0=\partial_+ \phi.
\end{equation}
The field $\nu$ is defined by
\begin{equation}
\label{43}
\partial_+ \nu= {\cosh 2r \over \cosh^2 r} \partial_+ \theta,
\quad
\partial_- \nu = \cosh^{-2} (r) \partial_- \theta
\end{equation}
Note that it follows from the field equations that
$$\partial_- ((\cosh 2r/ \cosh^2 r) \partial_+ \theta)
=\partial_+(\cosh^{-2} (r) \partial_- \theta).$$
Thus Eqs.\ref{43}  may be integrated. Using these relations, it
is easy to get convinced that $\partial_- W_{\pm}=0$,
$\partial_- W_0=0$, so that they define three conserved quantities
(characteristic integrals). One of them gives back the
stress energy tensor as expected, since $\theta_{++}= 4W_0$,
where $\theta_{\mu \nu}$ is given by Eq.\ref{2}.
In the same way as for the
$A_N$ abelian Toda case\cite{GM}, we expect that the
 other conserved
quantities will correspond to particular diffeomorphisms of
the target space.

Another general point is that there exists\cite{LSS}  a
 supersymmetrized version of the
dynamical system Eq.\ref{9}. In this, Eq.\ref{20} is replaced by
\begin{equation}
\label{44}
{\cal D}_+\left ( {\cal D}_-  \hat g_0^{-1} \hat g_0 \right )
+\left [ N_- , \hat  g_0^{-1} N_+ \hat g_0\right ]_+ =0.
\end{equation}
Here one uses a classical Lie super-algebra supplied with a
$\bf Z$-gradation. ${\cal D}_\pm$ are covariant super-derivatives in
$2|2$-super-space; $\hat g_0$ is a function on the
Grassmannian hull of the super-Lie-group generated by   ${\cal G}_0$.
$N_\pm$ are fixed (odd) elements of ${\cal G}_{\pm 1}$.
This gives a supersymmetrized black hole.
\bigskip

In conclusion, this discussion was at the purely classical level,
and the quantum case is left for further studies. We expect that
the complete integrability will be of great help to quantize
keeping conformal invariance, as it was already the case
for the abelian Toda systems.

\bigskip
\noindent{\large \bf Acknowledgements}

One of the authors (M. S.) is indebted to
 A. Leznov, Yu.~Manin,
A.~Razumov, and P.~Sorba for the useful discussions.
He is also grateful  to ENS LAPP in Lyon and LPTENS in Paris for
kind hospitality. This work was partially supported by the European
Community Twinning Program.

\end{document}